\documentclass[article,nojss]{jss}
\usepackage[utf8]{inputenc}
\providecommand{\tightlist}{%
  \setlength{\itemsep}{0pt}\setlength{\parskip}{0pt}}

\author{
Jacob Bien\\Departments of BSCB and Statistical Science, Cornell University
}
\title{The \pkg{simulator}: An Engine to Streamline Simulations}

\Abstract{
The \pkg{simulator} is an R package that streamlines the process of
performing simulations by creating a common infrastructure that can be
easily used and reused across projects. Methodological statisticians
routinely write simulations to compare their methods to preexisting
ones. While developing ideas, there is a temptation to write ``quick and
dirty'' simulations to try out ideas. This approach of rapid prototyping
is useful but can sometimes backfire if bugs are introduced. Using the
\pkg{simulator} allows one to remove the ``dirty'' without sacrificing
the ``quick.'' Coding is quick because the statistician focuses
exclusively on those aspects of the simulation that are specific to the
particular paper being written. Code written with the \pkg{simulator} is
succinct, highly readable, and easily shared with others. The modular
nature of simulations written with the \pkg{simulator} promotes code
reusability, which saves time and facilitates reproducibility. The
syntax of the \pkg{simulator} leads to simulation code that is easily
human-readable. Other benefits of using the \pkg{simulator} include the
ability to ``step in'' to a simulation and change one aspect without
having to rerun the entire simulation from scratch, the straightforward
integration of parallel computing into simulations, and the ability to
rapidly generate plots, tables, and reports with minimal effort.
}

\Plainauthor{Jacob Bien}
\Plaintitle{The Simulator: An Engine to Streamline Simulations}
\Shorttitle{The \pkg{simulator}}
\Plainkeywords{simulations, reproducible code, parallel computing}

\Submitdate{}

\Address{
    Jacob Bien\\
  Cornell University\\
  1178 Comstock Hall Cornell University Ithaca, NY 14853\\
  E-mail: \href{mailto:jbien@cornell.edu}{\nolinkurl{jbien@cornell.edu}}\\
  URL: \url{http://faculty.bscb.cornell.edu/~bien/}\\~\\
  }

\usepackage{amsmath, amsfonts}

\begin{document}

\newcommand{\real}{\mathbb R}

\section{Introduction}\label{introduction}

Methodological statisticians spend an appreciable amount of time writing
code for simulation studies. Every paper introducing a new method has a
simulation section in which the new method is compared across several
metrics to preexisting methods under various scenarios. Given the
formulaic nature of the simulation studies in most statistics papers,
there is a lot of code that can be reused. \textbf{The goal of the
\pkg{simulator} is to streamline the process of performing simulations
by creating a common infrastructure that can be easily used and reused.}
By infrastructure, we mean everything that is common across projects.
This includes code for

\begin{itemize}
\tightlist
\item
  running simulations in parallel,
\item
  proper handling of pseudorandom number generator streams (relevant
  when running simulations in parallel),
\item
  storing simulation outputs at various stages to allow one to ``step
  in'' and change one aspect of a simulation (such as how a method is
  being evaluated) without having to rerun the entire simulation from
  scratch,
\item
  summarizing simulation results with plots and tables showing how
  various methods compare across one or more metrics, and
\item
  generating reports to easily communicate simulation results and code
  to others.
\end{itemize}

The \pkg{simulator} is an R \citep{R} package focused on all of these
aspects of a simulation. This allows the statistician to focus
exclusively on problem-specific coding. The benefits of working this way
include

\begin{itemize}
\tightlist
\item
  a decrease in code size, which both reduces the amount of time spent
  coding and the likelihood of introducing bugs,
\item
  code that is easy to understand, share, and reuse,
\item
  the ability to maintain focus on the statistical issues at hand
  without either getting sidetracked by coding-related issues or
  alternatively settling for ``quick and dirty'' approaches that may
  save time in the present but slow down a project in the future (if one
  must, for example, rerun all results from scratch),
\item
  the ability to add to or modify a simulation without rerunning
  everything
\item
  the ability to open/close the R session without having to save the
  workspace (a common practice that produces unwieldy R environments
  that can lead to bugs in which global variables unintentionally affect
  behavior of functions)
\item
  a consistent organization of and approach to simulations across all
  projects, making it easier to resume a project that may have been
  dormant for years,
\item
  the ability to easily ``plug in'' components written by others for
  similar problems. Indeed, if a community of researchers (either at the
  research group level or at the level of a sub-area of statistics) is
  using the \pkg{simulator}, standard libraries of models, methods, and
  metrics can be developed at the community level, leading to easy code
  sharing.
\end{itemize}

The \pkg{simulator} divides a statistical simulation study into four
basic components or modules:

\begin{enumerate}
\def\labelenumi{\arabic{enumi})}
\item
  \textbf{Models:} The statistical model, determining how data is
  generated.
\item
  \textbf{Methods:} The statistical procedures being compared: Given the
  data, each method produces an output such as an estimate, prediction,
  or decision.
\item
  \textbf{Evaluations:} Metrics that evaluate a method's output based on
  the input.
\item
  \textbf{Plots, Tables, Reports:} How one displays the evaluated
  metrics for various methods under various scenarios, and how one
  communicates the results and code to others.
\end{enumerate}

When using the \pkg{simulator}, one codes the models, methods, and
metrics particular to the problem at hand. These are then ``plugged in''
to the \pkg{simulator}.

After an initial look at the \pkg{simulator} in action (Section
\ref{sec:first-look}), we will then discuss each of the components of
the \pkg{simulator} mentioned above in greater detail (Section
\ref{sec:components}). In Section \ref{sec:design}, we discuss certain
principles that informed our design choices for the package. Section
\ref{sec:getting-started} recommends a simple way to get started with
the \pkg{simulator}. Section \ref{sec:dependence-on-others} gives credit
to the R packages that are leveraged by the \pkg{simulator}, and Section
\ref{sec:related} reviews other simulation packages. We conclude (in
Section \ref{discussion}) with a discussion of the future of the
\pkg{simulator}.

\section{A first look: Betting on sparsity with the
lasso}\label{a-first-look-betting-on-sparsity-with-the-lasso}

\label{sec:first-look}

The easiest way to learn about the \pkg{simulator} is by example. For
this reason, we have created a series of vignettes, available online,
which show how the \pkg{simulator} can be used in the context of some of
the most well-known papers in statistics. We provide here a first look
at working with the \pkg{simulator}, applying it to a simulation
involving the lasso (an extended form of this example is available
online as a vignette).

\citet{ESL} put forward the ``bet on sparsity'' principle: \emph{``Use a
procedure that does well in sparse problems, since no procedure does
well in dense problems.''}

The authors perform simulations comparing the lasso \citep{Tibshirani96}
with ridge regression in sparse and dense situations. A simulation of
this sort can be concisely coded in several easy-to-read lines of code
using the \pkg{simulator}.

\begin{CodeChunk}
\begin{CodeInput}
library(simulator)
\end{CodeInput}
\end{CodeChunk}

\begin{CodeChunk}
\begin{CodeInput}
new_simulation(name = "bet-on-sparsity", label = "Bet on sparsity") %>%
  generate_model(make_sparse_linear_model, n = 200, p = 500, 
                 k = as.list(seq(5, 80, by = 5)), vary_along = "k") %>%
  simulate_from_model(nsim = 5, index = 1) %>%
  run_method(list(lasso, ridge)) %>%
  evaluate(list(mse, bestmse, df))
\end{CodeInput}
\end{CodeChunk}

The code above can be read as follows: Create a new simulation (with
some name and label); add to it a sequence of sparse linear models where
the sparsity level is varied from 1 to 80; simulate 5 random draws from
this model; run the lasso and ridge on both of these; finally, compute
the mean squared error, best mean squared error (over all values of the
tuning parameter), and degrees of freedom for each of these. In coding
this simulation, we only needed to define the problem-specific parts,
which are

\begin{enumerate}
\def\labelenumi{\arabic{enumi})}
\tightlist
\item
  the function \texttt{make\_sparse\_linear\_model}, which creates a
  \texttt{Model} object,
\item
  the \texttt{Method} objects \texttt{lasso} and \texttt{ridge},
\item
  and the \texttt{Metric} objects \texttt{mse}, \texttt{df}, and
  \texttt{bestmse}.
\end{enumerate}

This problem-specific code is provided in the Appendix.

The functions \texttt{new\_simulation}, \texttt{generate\_model},
\texttt{simulate\_from\_model}, \texttt{run\_method}, and
\texttt{evaluate} are \pkg{simulator} functions that take care of the
details of bookkeeping, random number generation, parallel processing
(if desired), saving outputs to file, etc.

\begin{figure}
\includegraphics[width=1.0\linewidth]{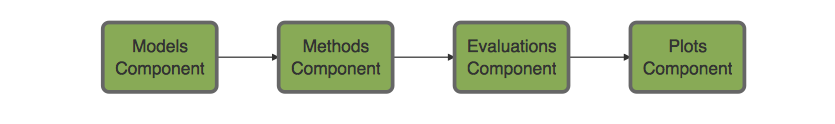}
\caption{A simulation can be viewed as a pipeline of interlocking components.}
\label{fig:flowchart}
\end{figure}

It is useful to think of a simulation as a pipeline (see Figure
\ref{fig:flowchart}), and this is emphasized by the \pkg{simulator}'s
use of the \texttt{magrittr} pipe \texttt{\%\textgreater{}\%}
\citep{magrittr}. We discuss the pipe further in Section
\ref{sec:magrittr}.

The results from all intermediate stages are saved to file. To get
access to these results for further analysis, one loads the
\texttt{Simulation} object that was created, using the name it was
given.

\begin{CodeChunk}
\begin{CodeInput}
sim <- load_simulation("bet-on-sparsity")
\end{CodeInput}
\end{CodeChunk}

This object only contains references to the simulation results (and not
the results themselves) and thus is quick to load and remains
lightweight even for large simulations.

We next make a plot showing how the best MSE of each method varies with
sparsity level.

\begin{CodeChunk}
\begin{CodeInput}
plot_eval_by(sim, "best_sqr_err", varying = "k", main = "Betting on sparsity")
\end{CodeInput}


\begin{center}\includegraphics{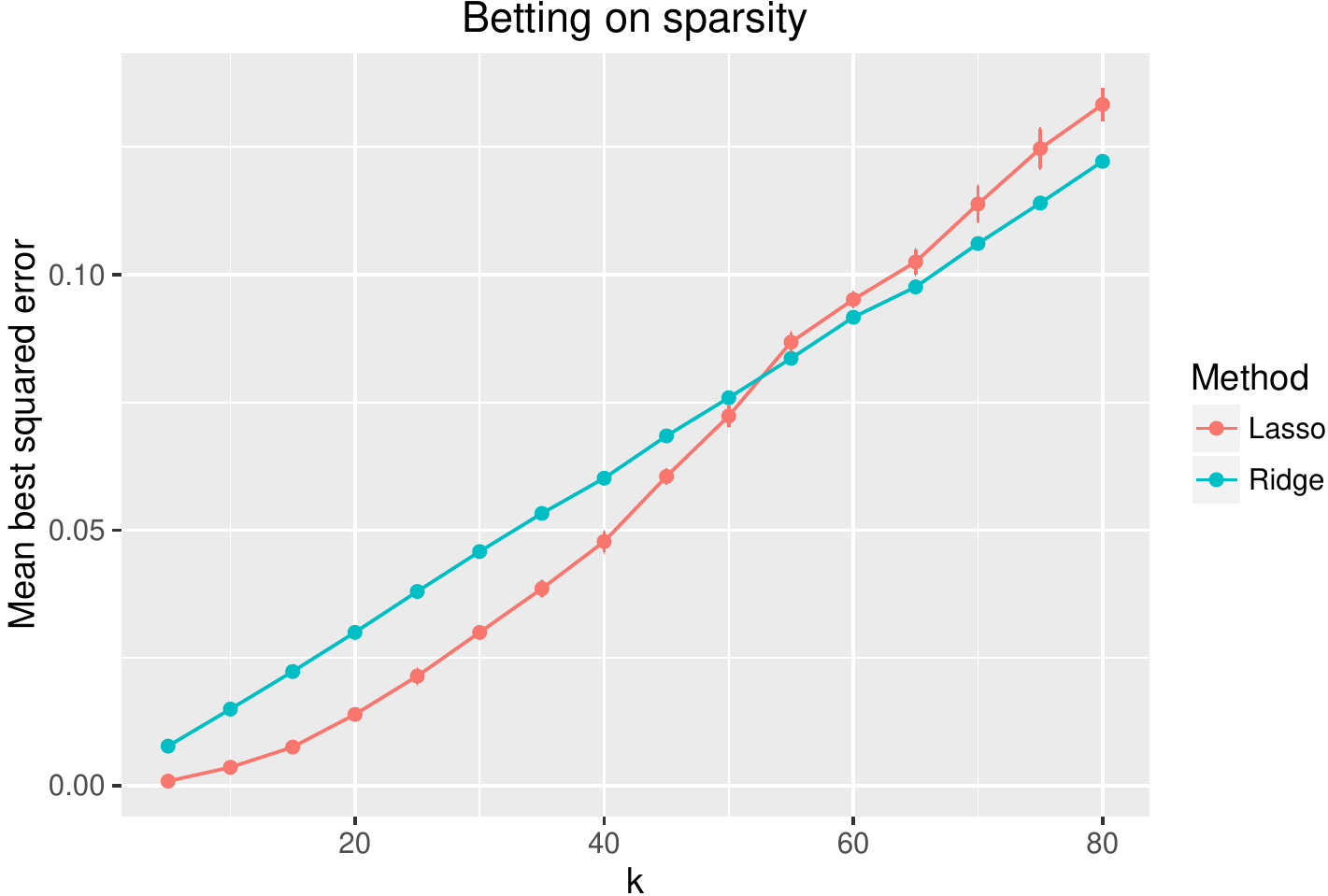} \end{center}

\end{CodeChunk}

The plot shows (a Monte Carlo estimate of) the best-achievable mean
squared error of both methods. The rationale for the ``bet on sparsity''
principle is apparent from the plot. When \(k\) is low, we see large
gains in performance using the lasso compared to ridge; when \(k\) is
large, ridge does better---however, in this regime, neither method is
doing particularly well (and in a relative sense, ridge's advantage is
only slight). Observe that the call to \texttt{plot\_eval\_by} is fairly
simple considering that the \pkg{simulator} has labeled the axes and has
identified the methods in a legend. This is an example of how the
formulaic nature of most simulation studies allows for efficiency in
coding.

The \pkg{simulator} is designed to make simulations as modular as
possible so that one can easily add new parts, run various steps in
parallel, etc. For example, if one later decides to compare to the
minimax concave penalty (MCP) \citep{zhang2010nearly}, one can simply
call

\begin{CodeChunk}
\begin{CodeInput}
sim <- run_method(sim, mcp)
\end{CodeInput}
\end{CodeChunk}

and the earlier steps of generating data will not be repeated. If one
decides to add more than 5 random draws to the simulation, one can do so
without having to redo the earlier computation (this will be discussed
in greater detail in Section \ref{sec:parallel}).

Rather than looking at just the best MSE of each method, we might also
like to examine how the MSE varies with degrees of freedom for each
method. We look at the methods' results when the true sparsity \(k\) is
low and when it is high.

\begin{CodeChunk}
\begin{CodeInput}
subset_simulation(sim, k == 20 | k == 80) %>% plot_evals("df", "sqr_err")
\end{CodeInput}


\begin{center}\includegraphics{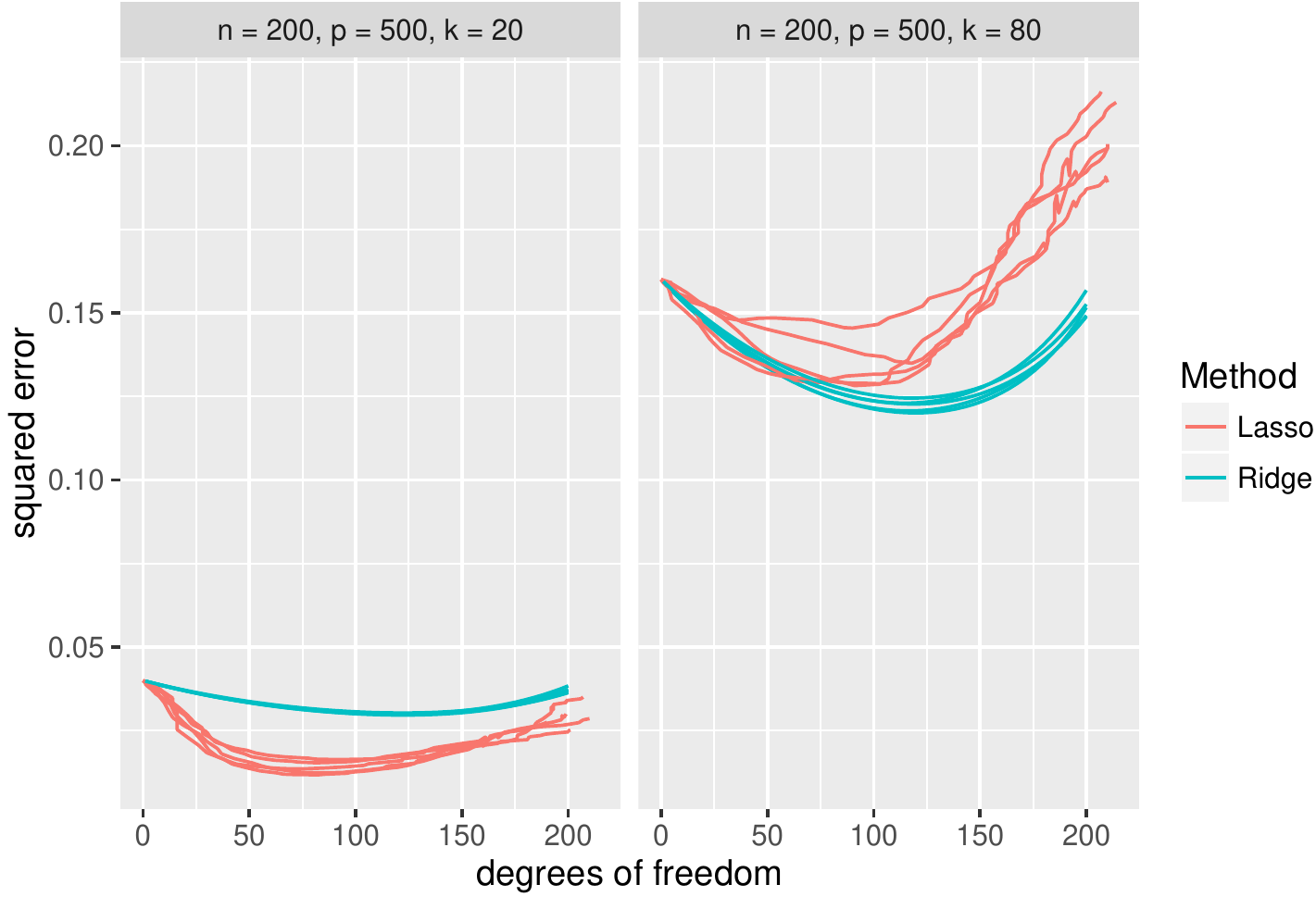} \end{center}

\end{CodeChunk}

In practice, it is common to use cross-validation to select the tuning
parameter, so we might also want to simulate the two methods when
cross-validation is used. We organize this as a new simulation that
shares the same models and simulated data as before, but with two new
methods.

\begin{CodeChunk}
\begin{CodeInput}
sim2 <- subset_simulation(sim, methods = "") %>%
  rename("bet-on-sparsity-cv") %>%
  relabel("Bet on sparsity (with cross validation)") %>%
  run_method(list(lasso + cv, ridge + cv)) %>%
  evaluate(mse)
\end{CodeInput}
\end{CodeChunk}

We explain in the Appendix in greater depth the \pkg{simulator}
construct that allows us to write \texttt{lasso\ +\ cv}, but for now we
note that doing so allows us to avoid rerunning the lasso and ridge from
scratch, which would be wasteful given that we have already computed
them in these situations in the previous simulation. A plot reveals that
qualitatively the results are quite similar.

\begin{CodeChunk}
\begin{CodeInput}
plot_eval_by(sim2, "sqr_err", varying = "k", main = "Betting on sparsity")
\end{CodeInput}


\begin{center}\includegraphics{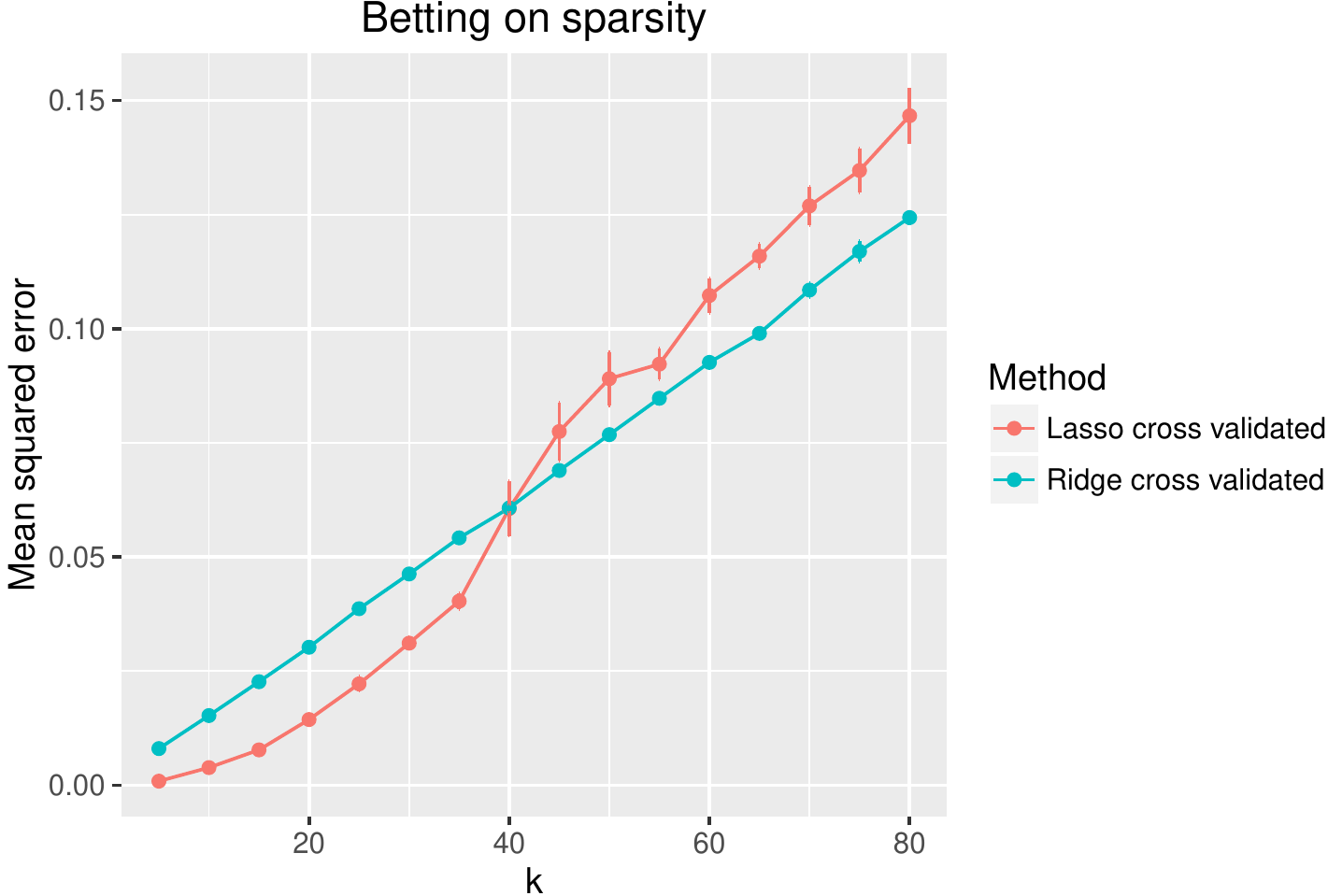} \end{center}

\end{CodeChunk}

\section{A look at each component}\label{a-look-at-each-component}

\label{sec:components}

In this section, we look more closely at the components of the
\pkg{simulator} and give brief overviews of the functions pertaining to
each component.

Thinking of a simulation as a pipeline, a \texttt{Simulation} object is
what gets pushed through the series of pipes. Each of the main pipeline
functions of the \pkg{simulator} (which can be thought of as the pipes)
input and output a simulation object. Each main pipeline function
creates new objects, saves them to file, and then records what was done
in the \texttt{Simulation} object. Thus, the \texttt{Simulation} object
serves as the memory of what has been done and as a handle for getting
to the various intermediate stages of the simulation. Importantly, a
\texttt{Simulator} object does not contain the results themselves, only
references to them. This is discussed in greater detail in Section
\ref{sec:design}.

\begin{itemize}
\tightlist
\item
  \texttt{new\_simulation} is used to create a new \texttt{Simulation}
  object. By default, when it is called, a record of its existence is
  saved to file.
\item
  \texttt{save\_simulation} is passed a \texttt{Simulation} object, it
  saves it to file. The location is given by the \texttt{name} and
  \texttt{dir} (short for directory) slot of the object.
\item
  \texttt{load\_simulation} when passed the name of a simulation loads
  the \texttt{Simulation} object into the R session.
\item
  \texttt{subset\_simulation} is useful for creating a new
  \texttt{Simulation} object from a preexisting one. For example, this
  is useful for focusing a plot on only certain models or for reusing
  only certain simulation results in a different context.
\end{itemize}

\subsection{Models component}\label{models-component}

The first component in the \pkg{simulator} is the models component. An
object of class \texttt{Model} defines a set of parameters and a
function, called \texttt{simulate}, that takes these parameters and
returns a specified number of random draws from this distribution. For
example, in the ``bet on sparsity'' example of Section
\ref{sec:first-look}, we created a sequence of \texttt{Model} objects
with varying sparsity level \texttt{k}. Loading one such object, we can
have a look by printing it:

\begin{CodeChunk}
\begin{CodeInput}
m <- model(sim, k == 10)
m
\end{CodeInput}
\begin{CodeOutput}
Model Component
 name: slm/k_10/n_200/p_500
 label: n = 200, p = 500, k = 10
 params: x beta mu sigma n p k
\end{CodeOutput}
\end{CodeChunk}

The parameters are stored in a named list \texttt{m@params} and the
function to simulate data is \texttt{m@simulate}.

Here are some of the most useful functions related to the model
component:

\begin{itemize}
\tightlist
\item
  \texttt{new\_model} creates a new \texttt{Model} object. In the
  example, it is called in \texttt{make\_sparse\_linear\_model} (see
  Appendix \ref{sec:lasso-model}).
\item
  \texttt{generate\_model} is one of the main pipeline functions of the
  \pkg{simulator}. The user supplies it with a function
  (\texttt{make\_sparse\_linear\_model}, in the example) that returns an
  object of class \texttt{Model}, and \texttt{generate\_model} then
  creates this model, saves it to file, and adds a reference to it to
  the \texttt{Simulation} object. In many simulation studies, one is
  interested in a sequence of models, indexed by several parameters. The
  \texttt{vary\_along} parameter of \texttt{generate\_model} is used to
  specify which parameters should be varied.
\item
  \texttt{simulate\_from\_model} is another main pipeline function. Once
  the model has been generated, \texttt{simulate\_from\_model} creates a
  specified number of random draws from the distribution defined by the
  model. The \texttt{Draws} objects are saved to file and references to
  them are added to the \texttt{Simulation} object. These simulations
  can be broken into chunks and run either sequentially or in parallel
  (or a mix of the two). Parallelization is described in greater detail
  in Section \ref{sec:parallel}.
\item
  \texttt{model} loads one or more \texttt{Model} objects from a
  \texttt{Simulation} object into the R session.
\item
  \texttt{draws} loads one or more \texttt{Draws} objects from a
  \texttt{Simulation} object into the R session.
\end{itemize}

The \texttt{Draws} objects have a simple structure.

\begin{CodeChunk}
\begin{CodeInput}
d <- draws(sim, k == 10)
d
\end{CodeInput}
\begin{CodeOutput}
Draws Component
 name: slm/k_10/n_200/p_500
 label: (Block 1:) 5 draws from n = 200, p = 500, k = 10
\end{CodeOutput}
\end{CodeChunk}

In this example, \texttt{d@draws} is a list of length 5 (since we had
\texttt{nsim\ =\ 5}), with each element being a realization of the
response vector \(y\).

\subsection{Methods component}\label{methods-component}

The next component in the \pkg{simulator} is the methods component. An
object of class \texttt{Method} defines a function that takes arguments
\texttt{model} and \texttt{draw}. In the example, the design matrix
\(X\) is contained in the model (accessible via \texttt{model\$x}), and
the response vector \(y\) comes from the draw argument. A
\texttt{Method} object can also (optionally) include a list of
\texttt{settings}.

\begin{itemize}
\tightlist
\item
  \texttt{new\_method} creates a new \texttt{Method} object. In the
  example, the objects \texttt{lasso} and \texttt{ridge} were produced
  by the \texttt{new\_method} function (see Appendix
  \ref{sec:lasso-methods}).
\item
  \texttt{run\_method} is a main pipeline function. It runs one or more
  methods on the simulated data and creates \texttt{Output} objects,
  saves these to file, and adds references to these to the
  \texttt{Simulation} object.
\item
  \texttt{output} loads one or more \texttt{Output} objects from a
  \texttt{Simulation} object into the R session.
\item
  (for advanced usage---not needed in most simulations)
  \texttt{new\_method\_extension} and \texttt{new\_extended\_method} are
  used to create new \texttt{MethodExtension} and
  \texttt{ExtendedMethod} objects. An \texttt{ExtendedMethod} is an
  object that behaves like a method, but is allowed to use the output of
  another method. This can be useful when one method builds off of
  another. In the example, \texttt{cv} is a \texttt{MethodExtension} and
  \texttt{lasso\ +\ cv} is an \texttt{ExtendedMethod} (see Appendix
  \ref{sec:method-extensions} for more).
\end{itemize}

At the core of an \texttt{Output} object is a list called \texttt{out}.

\begin{CodeChunk}
\begin{CodeInput}
o <- output(sim, k == 10, methods = "lasso")
o
\end{CodeInput}
\begin{CodeOutput}
Output Component
 model_name: slm/k_10/n_200/p_500
 index: 1
 nsim: 5
 method_name: lasso
 method_label: Lasso
 out: beta, yhat, lambda, df, time
\end{CodeOutput}
\end{CodeChunk}

Each element of the list \texttt{o@out} corresponds to a random draw
(i.e., realization) and gives the output of the lasso on this draw.

\subsection{Metrics/Evaluation
component}\label{metricsevaluation-component}

The next component in the \pkg{simulator} is the evaluations component,
in which we evaluate each method based on some user-defined metrics. An
object of class \texttt{Metric} defines a function with arguments
\texttt{model} and \texttt{out} (the output of a method). In the
example, the mean squared error is a metric of interest which measures
how far the true coefficient vector \(\beta\) (accessible via
\texttt{model\$beta}) is from a method's estimate of this quantity
(accessible via \texttt{out\$beta}).

\begin{itemize}
\tightlist
\item
  \texttt{new\_metric} creates a new \texttt{Metric} object. In the
  example, the objects \texttt{mse}, \texttt{bestmse}, and \texttt{df}
  are produced by the \texttt{new\_method} function (see Appendix
  \ref{sec:lasso-metrics}).
\item
  \texttt{evaluate} is a main pipeline function. It computes one or more
  metrics on the outputs of the methods and creates \texttt{Evals}
  objects, saves these to file, and adds references to these to the
  \texttt{Simulation} object. In addition to the metrics that are
  explicitly passed to \texttt{evaluate}, the computing time for a
  method is automatically saved as well.
\item
  \texttt{evals} loads one or more \texttt{Evals} objects from a
  \texttt{Simulation} object into the R session.
\item
  \texttt{as.data.frame} converts an \texttt{Evals} object (or list of
  \texttt{Evals} objects) to a \texttt{data.frame}.
\end{itemize}

An \texttt{Evals} object contains the computed metrics on each
method-draw pair.

\begin{CodeChunk}
\begin{CodeInput}
e <- evals(sim, k == 10)
e
\end{CodeInput}
\begin{CodeOutput}
Evals Component
 model_name: slm/k_10/n_200/p_500    index: 1 (5 nsim)
 method_name(s): lasso, ridge (labeled: Lasso, Ridge)
 metric_name(s): sqr_err, df, best_sqr_err, time
 metric_label(s): squared error, degrees of freedom, best squared error, Computing time
\end{CodeOutput}
\end{CodeChunk}

In particular, \texttt{e@evals\$lasso} and \texttt{e@evals\$ridge} are
lists of length 5 (one per random draw) and
\texttt{e@evals\$lasso{[}{[}1{]}{]}\$bestmse} gives the best MSE for the
lasso on the first draw.

\subsection{Plot/Table component}\label{plottable-component}

Making standard plots and tables of simulation results is a task that is
particularly streamlined using the \pkg{simulator}. -
\texttt{plot\_eval} takes the name of a metric and makes side-by-side
box plots of each method. If multiple models are included, these are
shown as a multifaceted plot.

\begin{CodeChunk}
\begin{CodeInput}
subset_simulation(sim, k == 20 | k == 80) %>% plot_eval("time")
\end{CodeInput}


\begin{center}\includegraphics{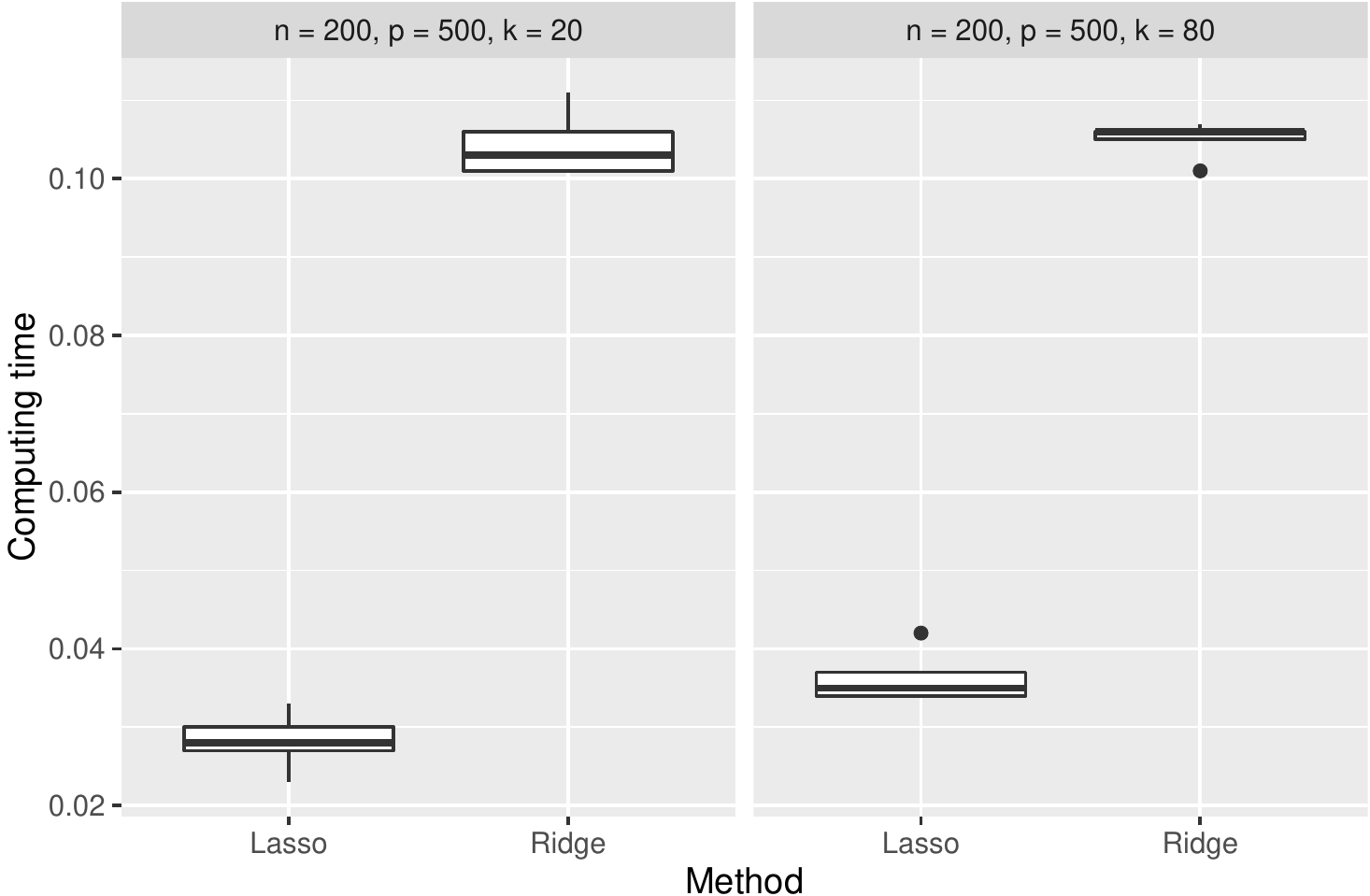} \end{center}

\end{CodeChunk}

\begin{itemize}
\tightlist
\item
  \texttt{plot\_evals} takes the names of two metrics and plots the
  first versus the second with different colored/styled lines (or
  points) for each method. We used this function in Section
  \ref{sec:first-look}.
\item
  \texttt{plot\_eval\_by} plots a metric versus a model parameter that
  was varied across models (in our example, this corresponded to the
  sparsity level \texttt{k} of the true coefficent vector). By default,
  a line with error bars is shown. The center and width of these error
  bars is by default the sample mean (across the random realizations)
  and standard error of the sample mean, respectively; however, one can
  pass custom-made \texttt{Aggregator} objects to have the center and
  width of the error bars represent different aggregates. Another option
  (with \texttt{type="raw"}) simply shows the raw values of the metric
  for each method at each value of the model parameter.
\item
  \texttt{tabulate\_eval} takes the name of a metric and creates a table
  (in latex, markdown, or html formats) in which each row is a model and
  each column is a method. As with \texttt{plot\_eval\_by}, the default
  is to show the sample mean (with its standard error in parentheses),
  and the user can provide custom \texttt{Aggregator} objects so that
  these represent more general aggregates.
\item
  (for advanced usage---not needed in most simulations)
  \texttt{new\_aggregator} can be used to create a custom
  \texttt{Aggregator}.
\end{itemize}

\section{Design principles}\label{design-principles}

\label{sec:design}

\subsection{Modularity of components}\label{modularity-of-components}

As noted earlier, a simulation can be viewed as a pipeline of
interlocking components. Each component can be easily inserted,
modified, or removed without requiring one to change the structure of
the overall simulation. For example, if one wishes to consider an
additional model, include another method, or evaluate the methods using
a different metric, this can all be easily done with minimal structural
changes to the overall simulation. Each component in the pipeline has a
standardized input and is responsible for providing output that will fit
into the next component's input. For example, at the heart of a
\texttt{Metric} object is a function that takes in a method's output and
a \texttt{Model} and outputs a number or vector of numbers. The output
of a metric can then be easily fed into the \texttt{simulator}'s plot or
tabulation functions.

Another aspect of modularity is that the name (a short identifier used
in file names) and the label (a longer human-readable string) of the
component are stored as part of the component. This is in contrast to
the typical practice of only labeling methods or models when plots or
tables are made. Conceptually, it makes much more sense to label
components at the location they are defined.\footnote{This is similar in
  spirit to \pkg{roxygen2} \citep{roxygen2}, in which one documents
  functions in the same place where they are defined rather than writing
  separate \texttt{.Rd} files.} A great benefit is that all downstream
components have access to the labels. For example, the \pkg{simulator}'s
plot functions automatically provide informative descriptions such as
legends identifying the various methods, labeled axes, and titles
describing the model.

Decoupling the components of a simulation is also important when certain
aspects require more computing time than others. For example, applying
statistical methods to simulated data is often much slower than
computing metrics to evaluate their success. The modularity of the
\pkg{simulator} disentangles these two parts so that if one changes a
single metric, one does not have to rerun the statistical methods.
Finally, the plotting and tabulating components are themselves separate
from all previous parts---since one should not have to rerun any major
computation as one makes slight modifications to plots or tables.

Finally, modularity makes it easier to share components across one's
projects with other researchers. A group of researchers might develop a
set of common models and methods for, say, high-dimensional covariance
estimation. When someone proposes a new covariance estimator, one simply
needs to define this new \texttt{Method} object and then this can be
easily plugged in to the preexisting code base, saving a lot of time and
making for a more meaningful comparison.

\subsection{Every component's results saved to
file}\label{every-components-results-saved-to-file}

Every component of a simulation generates R objects that take time to
compute and take space to keep in memory. When we modify a component of
a simulation, we should not have to rerun the parts of the simulation
that are ``upstream'' of the modification. Also, in analyzing a
simulation, we might want to examine the generated objects at various
stages in the simulation; again, we should not have to rerun anything to
see an intermediate stage of the simulation. For all of these reasons,
the \pkg{simulator} automatically saves all generated objects to files
when one runs a simulation. The files are kept in an organized directory
structure that is easy to interpret; however, the user never has to
explicitly learn or pay attention to the particulars of the directory
structure since there are a series of \pkg{simulator} functions that
make it simple to load these saved files.

\subsection{Working with references}\label{working-with-references}

In the most common usage of the \pkg{simulator}, a \texttt{Simulation}
object is passed through the \pkg{simulator} pipeline. As it gets fed
through the various component functions, it accumulates more and more of
a ``record'' of the simulation. An important aspect is that the
\texttt{Simulation} object does not contain the objects generated from
the various components themselves (which would be extremely memory
intensive); rather, it contains \emph{references} to these objects. A
reference is an object containing the saved location of an object
generated by the \pkg{simulator} The fact that \texttt{Simulation}
objects contain references rather than the objects themselves makes the
\pkg{simulator} behave much more nimbly. One only loads the specific
pieces of a simulation as needed. The functions \texttt{models},
\texttt{draws}, \texttt{outputs}, and \texttt{evals} allow one to load
specific objects referred to in a simulation. A more advanced
consequence is that multiple simulations can refer to the same set of
saved objects (thus if multiple simulations have some common aspects,
one can avoid recomputing these or creating multiple copies of them).
The function \texttt{subset\_simulation} is useful in this context.

\subsection{Magrittr friendly}\label{magrittr-friendly}

\label{sec:magrittr}

The first argument and the output of most \pkg{simulator} functions is a
\texttt{Simulation} object. This is done to facilitate \pkg{magrittr}
integration, which allows one to write succinct ``one-liner''
simulations that read (almost) like sentences. The pipe operator passes
the output of its left-hand-side to the first argument of the function
on its right.\footnote{For example,
  \texttt{matrix(rnorm(5\ *\ 2),\ 5,\ 2)} could equivalently be written
  \texttt{5\ *\ 2\ \%\textgreater{}\%\ rnorm\ \%\textgreater{}\%\ matrix(5,\ 2)}.}
Without the pipe, the example presented in Section \ref{sec:first-look}
would be a bit more cumbersome.

\begin{CodeChunk}
\begin{CodeInput}
sim <- new_simulation(name = "bet-on-sparsity", label = "Bet on sparsity")
sim <- generate_model(sim, make_sparse_linear_model, n = 200, p = 500, 
                 k = as.list(seq(5, 80, by = 5)), vary_along = "k")
sim <- simulate_from_model(sim, nsim = 5, index = 1)
sim <- run_method(sim, list(lasso, ridge))
sim <- evaluate(sim, list(mse, bestmse, df))
\end{CodeInput}
\end{CodeChunk}

\subsection{Parallel processing and random seed
streams}\label{parallel-processing-and-random-seed-streams}

\label{sec:parallel}

The \pkg{simulator} uses the \pkg{parallel} package \citep{R} to allow
simulations to be run in parallel. In \texttt{simulate\_from\_model} one
simulates draws in chunks, indexed by \texttt{index}. For example,
\texttt{simulate\_from\_model(sim,\ nsim\ =\ 10,\ index\ =\ 1:10)} would
lead to a total of 100 simulations performed in chunks of size 10. The
\texttt{index} of a chunk is used to specify a distinct stream of
pseudorandom numbers, using the ``L'Ecuyer-CMRG'' generator in R
\citep{lecuyer}. The use of streams is convenient because it
compartmentalizes chunks of random draws, so that the starting state of
one chunk does not depend on the end state of another chunk. In
particular, the following three options all give identical results.

\begin{enumerate}
\def\labelenumi{\arabic{enumi})}
\tightlist
\item
  In sequence:
\end{enumerate}

\begin{CodeChunk}
\begin{CodeInput}
simulate_from_model(sim, nsim = 10, index = 1:10) %>% 
  run_methods(list_of_methods)
\end{CodeInput}
\end{CodeChunk}

\begin{enumerate}
\def\labelenumi{\arabic{enumi})}
\setcounter{enumi}{1}
\tightlist
\item
  In parallel:
\end{enumerate}

\begin{CodeChunk}
\begin{CodeInput}
simulate_from_model(sim, nsim = 10, index = 1:10) %>%
  run_methods(list_of_methods, parallel = list(socket_names = 4))
\end{CodeInput}
\end{CodeChunk}

\begin{enumerate}
\def\labelenumi{\arabic{enumi})}
\setcounter{enumi}{2}
\tightlist
\item
  Mixed-and-matched with no particular order to \texttt{index}:
\end{enumerate}

\begin{CodeChunk}
\begin{CodeInput}
simulate_from_model(sim, nsim = 10, index = 1:2) %>% 
  run_methods(list_of_methods)
# perhaps a day later:
sim <- load_simulation("bet-on-sparsity")
simulate_from_model(sim, nsim = 10, index = 5:10) %>% 
  run_methods(list_of_methods, parallel = list(socket_names = 4))
# realize later you forgot two chunks:
sim <- load_simulation("bet-on-sparsity")
simulate_from_model(sim, nsim = 10, index = 3:4) %>%
  run_methods(list_of_methods)
\end{CodeInput}
\end{CodeChunk}

When \texttt{simulate\_from\_model} creates a new \texttt{Draws} object
(i.e., a chunk of random draws) the end state of the random number
generator is saved to file. Whenever \texttt{run\_method} is called on
that \texttt{Draws} object, the random number generator is first set to
that stored state. This ensures that we will get the same results
regardless of the order in which we apply our methods (if any are
randomized algorithms).

\subsection{Unified interface for plotting and
tables}\label{unified-interface-for-plotting-and-tables}

\citet{gelman2002} go through every table appearing in an issue of the
\emph{Journal of the American Statistical Association} and conclude that
plots are more effective than tables for most tasks. The \pkg{simulator}
therefore focuses mostly on the generation of plots rather than tables.
That said, the basic function for making tables (which can be in latex,
html, or markdown format) is designed to have a very similar interface
to one of the plotting functions, reflecting the similar task these two
functions accomplish. For example, returning to the comparison of lasso
and ridge with cross validation, one could tabulate the results (even
though the plot generated in Section \ref{sec:first-look} seems more
useful).

\begin{CodeChunk}
\begin{CodeInput}
subset_simulation(sim2, k > 30 & k <= 60) %>% 
  tabulate_eval("sqr_err", format = list(nsmall = 2, digits = 0))
\end{CodeInput}
\begin{table}

\caption{A comparison of Mean squared error (averaged over 5 replicates).}
\centering
\begin{tabular}[t]{l|l|l}
\hline
  & Lasso cross validated & Ridge cross validated\\
\hline
n = 200, p = 500, k = 35 & 0.04 (0.00) & 0.05 (0.00)\\
\hline
n = 200, p = 500, k = 40 & 0.06 (0.01) & 0.06 (0.00)\\
\hline
n = 200, p = 500, k = 45 & 0.08 (0.01) & 0.07 (0.00)\\
\hline
n = 200, p = 500, k = 50 & 0.09 (0.01) & 0.08 (0.00)\\
\hline
n = 200, p = 500, k = 55 & 0.09 (0.00) & 0.08 (0.00)\\
\hline
n = 200, p = 500, k = 60 & 0.11 (0.00) & 0.09 (0.00)\\
\hline
\end{tabular}
\end{table}

\end{CodeChunk}

Everything about the table created by this command, including the
caption, was automatically generated by \texttt{tabulate\_eval}.

\subsection{Easy extraction of data into
data.frames}\label{easy-extraction-of-data-into-data.frames}

The \pkg{simulator} is intended to provide all the major functionality
that is typically required when one performs a simulation study;
however, users should not feel ``locked in'' to the \pkg{simulator}. In
particular, since there may be specific situations in which a
non-standard operation is required (e.g., generating an unusual sort of
plot based on simulation results), the \pkg{simulator} makes it simple
to extract simulation data into a \texttt{data.frame} which can then be
manipulated outside the \pkg{simulator} framework.

\section{Getting started}\label{getting-started}

\label{sec:getting-started}

The easiest way to start using the \pkg{simulator} is to use the
function \texttt{create}. The command

\begin{CodeChunk}
\begin{CodeInput}
create("name/of/new/directory")
\end{CodeInput}
\end{CodeChunk}

will produce the specified directory and the following files that make
up the skeleton of a simulation:

\begin{itemize}
\item
  \texttt{model\_functions.R} - file with a function that creates
  \texttt{Model} objects.
\item
  \texttt{method\_functions.R} - file that creates \texttt{Method}
  objects.
\item
  \texttt{eval\_functions.R} - file that creates \texttt{Metric}
  objects.
\item
  \texttt{main.R} - file with the main simulation pipeline code (and
  that sources the three \texttt{*\_functions.R} files).
\item
  \texttt{writeup.Rmd} - an \texttt{rmarkdown} file that lays out the
  code and results (showing the main code first and the definitions of
  various components later). The simulation code is not itself stored in
  \texttt{writeup.Rmd}, so that as one makes changes to the \texttt{R}
  files, the report will remain up-to-date. Some basic caching logic is
  built into \texttt{writeup.Rmd} so that if the \texttt{.R} files of
  the simulation have not been changed since the simulation data was
  created, the simulation will not be rerun when one knits
  \texttt{writeup.Rmd}. This is convenient since one can see the effects
  of changes to the report without having to rerun the entire
  simulation.
\end{itemize}

A series of online vignettes for the \pkg{simulator} demonstrate its use
in varied settings and may be useful for getting started with the
\pkg{simulator}.

\section{Dependence on other R
packages}\label{dependence-on-other-r-packages}

\label{sec:dependence-on-others}

The \pkg{simulator} makes use of several \texttt{R} packages. The pipe
operation, \texttt{\%\textgreater{}\%} uses \pkg{magrittr}
\citep{magrittr}; the parallel processing capabilities are built upon
the \pkg{parallel} \citep{R} package; file name formation is based on
the \pkg{digest} package \citep{digest}; \pkg{knitr} \citep{knitr} and
\pkg{rmarkdown} \citep{rmarkdown} are used for creating tables and
report generation; plots are by default generated using \pkg{ggplot2}
\citep{ggplot2}, although each plot function has a
\texttt{use\_ggplot2\ =\ FALSE} setting for users preferring the basic
plot functions provided through \texttt{graphics}; finally, although not
directly involved in user-called functions, the packages \pkg{devtools}
\citep{devtools} and \pkg{testthat} \citep{testthat} were instrumental
in the development of the package.

\section{Related work}\label{related-work}

\label{sec:related}

A look into the literature and a search online for software packages
reveal several packages similar in spirit to the \pkg{simulator}, which
we briefly review here.

The packages \pkg{ezsim} \citep{ezsim} and \pkg{simsalapar}
\citep{simsalapar} are similar in spirit in many ways to the
\pkg{simulator} and provide much of the same functionality (such as the
ability to parallelize simulations, save results to files, generate
plots, and vary model parameters). The major differences are in terms of
design choices. For example, these packages' approaches are much less
modular and less object-oriented. All aspects of the simulation are
passed as functions to a single master function. By contrast, the
\pkg{simulator} design encourages one to think of a simulation as a
pipeline with replaceable parts that can be easily reused, swapped, and
shared. Another difference is that \pkg{ezsim} does not appear
well-suited for simulations that evolve over time (for example, if one
later decides to add a method, increase the number of random draws, or
add an additional metric, the \pkg{simulator} does not require one to
rerun everything).

\citet{alfons2010object} develop the package \pkg{simFrame} as an
object-oriented framework for simulation (and also include
parallelization and plotting abilities). Their particular focus is on
survey statistics (with associated issues of outliers and missing data),
but the authors note that the framework is general and can be extended
to many other applications. In the examples presented, it appears that
the user writes a function that is passed to
\texttt{simFrame::runSimulation} that has all the methods and metrics
computed within it. This bundling of all methods and metrics together
goes against the \pkg{simulator} design principle that methods should be
decoupled from each other (so they can be run in parallel or at
different times) and that the computation of metrics to evaluate the
method objects should be viewed as a separate layer of the simulation.

The package \pkg{harvestr} \citep{harvestr} focuses on the
parallelization and caching infrastructure (and adds color with function
names related to gardening!); the framework is quite general but less
tailored to the sort of statistical simulations in which one is
comparing methods across different models and wants to rapidly generate
plots.

\pkg{SimDesign} \citep{simDesign} has many aspects in common with the
\pkg{simulator} and \pkg{simFrame}, for example its pipeline of
generate-analyse-summarise is similar to the model-method-evaluate
pipeline of the \pkg{simulator}. However, like the other simulation
packages, \pkg{SimDesign}'s methods appear to be bundled together in the
analyse step. Likewise all metrics are bundled together in the summarize
step. Again, a distinguishing feature of the \pkg{simulator} is the
modular organization of the Models/Methods/Metrics components.

\section{Discussion}\label{discussion}

The \pkg{simulator} attempts to streamline an important yet often
dreaded part of writing statistics papers. By taking care of the
infrastructure and enforcing some structure on the process, we have
found that it makes conducting simulation studies less tedious and less
haphazard. Using this tool allows the user to focus time, energy, and
thought on the problem-specific aspects, which hopefully leads to more
carefully thought-out, insightful studies that might even be enjoyable
to conduct. The foundation provided by the \pkg{simulator} can easily be
built upon by developing domain-specific libraries of models, methods,
and metrics. For example, the areas of high-dimensional regression,
classification, multiple testing, graphical models, clustering, network
science, etc, could each have a repository of \pkg{simulator}
\texttt{Model}, \texttt{Method}, and \texttt{Metric} objects that can be
easily shared. Building up such a library of components would greatly
facilitate the sharing of simulation models, leading to greater
consistency across the literature while at the same time requiring less
effort.

\newpage
\section*{Acknowledgments}\label{acknowledgments}
The inspiration to take the large code base we had developed over years
and turn it into an easy-to-use \texttt{R} package came from
\pkg{devtools}, which was also used to make this package. We also
acknowledge the book ``R packages'' \citep{wickham2015r} for influencing
many of the design and code-style choices for the \pkg{simulator}
package.

\bibliography{refs.bib}
\appendix
\section{Appendix}\label{appendix}

This appendix provides the problem-specific code required for the
example shown in Section \ref{sec:first-look}.

\subsection{The model}\label{the-model}

\label{sec:lasso-model}

We simulate from a linear model \[
Y=X\beta + \epsilon
\] where \(Y\in\real^n\), \(\beta\in\real^p\), and
\(\epsilon\sim N(0,\sigma^2I_n)\). We have taken \(X\) to have iid
\(N(0,1)\) entries and treat it as fixed in this simulation. We define a
\texttt{Model} object, which specifies the parameters and, most
importantly, describes how to simulate data.

\begin{verbatim}
make_sparse_linear_model <- function(n, p, k) {
  x <- matrix(rnorm(n * p), n, p)
  beta <- rep(c(1, 0), c(k, p - k))
  mu <- as.numeric(x %*% beta)
  sigma <- sqrt(sum(mu^2) / (n * 2)) # fixes signal-to-noise at 2
  new_model(name = "slm", label = sprintf("n = %s, p = %s, k = %s", n, p, k),
            params = list(x = x, beta = beta, mu = mu, sigma = sigma, n = n,
                          p = p, k = k),
            simulate = function(mu, sigma, nsim) {
              y <- mu + sigma * matrix(rnorm(nsim * n), n, nsim)
              return(split(y, col(y))) # make each col its own list element
            })
}
\end{verbatim}

We will typically put the code above in a file named
\texttt{model\_functions.R}.

\subsection{The methods}\label{the-methods}

\label{sec:lasso-methods}

We compare the lasso and ridge. Both of these methods depend on tuning
parameters, so we compute a sequence of solutions.

\begin{verbatim}
library(glmnet)
lasso <- new_method("lasso", "Lasso",
                    method = function(model, draw, lambda = NULL) {
                      if (is.null(lambda))
                        fit <- glmnet(x = model$x, y = draw, nlambda = 50,
                                      intercept = FALSE)
                      else {
                        fit <- glmnet(x = model$x, y = draw, lambda = lambda,
                                      intercept = FALSE)
                      }
                      list(beta = fit$beta, yhat = model$x %*% fit$beta,
                           lambda = fit$lambda, df = fit$df)
                    })

ridge <- new_method("ridge", "Ridge",
                    method = function(model, draw, lambda = NULL) {
                      sv <- svd(model$x)
                      df_fun <- function(lam) {
                        # degrees of freedom when tuning param is lam
                        sum(sv$d^2 / (sv$d^2 + lam))
                      }
                      if (is.null(lambda)) {
                        nlambda <- 50
                        get_lam <- function(target_df) {
                          f <- function(lam) df_fun(lam) - target_df
                          uniroot(f, c(0, 100 * max(sv$d^2)))$root
                        }
                        lambda <- sapply(seq(1, nrow(model$x),
                                             length = nlambda), get_lam)
                      }
                      df <- sapply(lambda, df_fun)
                      beta <- sapply(lambda, function(r) {
                        d <- sv$d / (sv$d^2 + r)
                        return(sv$v %*% (d * crossprod(sv$u, draw)))
                      })
                      list(beta = beta, yhat = model$x %*% beta,
                           lambda = lambda, df = df)
                    })
\end{verbatim}

Methods can return different items. However, aspects of the method that
will be used downstream in the simulation and compared across methods
should be in a common format. Thus \texttt{beta}, \texttt{yhat}, and
\texttt{df} in each case are in the same format. These will be the items
used when evaluating the methods' performances.

We will typically put the code above in a file named
\texttt{method\_functions.R}.

\subsection{The metrics}\label{the-metrics}

\label{sec:lasso-metrics}

When we compare methods in plots and tables, there are usually a number
of ``metrics'' we use. An object of class \texttt{Metric} specifies how
to go from a model's parameters and the output of a method and return
some quantity of interest.

\begin{verbatim}
mse <- new_metric("mse", "Mean-squared error",
                  metric = function(model, out) {
                    colMeans(as.matrix(out$beta - model$beta)^2)
                  })

bestmse <- new_metric("bestmse", "Best mean-squared error",
                      metric = function(model, out) {
                        min(colMeans(as.matrix(out$beta - model$beta)^2))
                      })

df <- new_metric("df", "Degrees of freedom",
                 metric = function(model, out) out$df)
\end{verbatim}

Observe that \texttt{out} refers to the list returned by our methods and
\texttt{model} refers to the \texttt{Model} object that is generated by
\texttt{make\_sparse\_linear\_model}. The \texttt{\$} operator can be
used to get parameters that are stored in the \texttt{Model} object.

We will typically put the code above in a file named
\texttt{eval\_functions.R}.

\subsection{Extensions of the methods}\label{extensions-of-the-methods}

\label{sec:method-extensions}

In Section \ref{sec:first-look}, we wanted to study cross-validated
versions of the lasso and ridge regression, so we inputed into
\texttt{run\_method} the objects \texttt{lasso\ +\ cv} and
\texttt{ridge\ +\ cv}. These are \texttt{ExtendedMethod} objects, which
behave much like \texttt{Method} objects, except they can get access to
the output of another method.

If we only cared about relaxing the lasso, we could have directly
created an \texttt{ExtendedMethod}; however, in this situation we wanted
both methods to be cross-validated in an identical fashion. In the
spirit of code reusability, we therefore created what we call a
\texttt{MethodExtension} object, \texttt{cv}. A \texttt{MethodExtension}
object when added to a \texttt{Method} object generates an
\texttt{ExtendedMethod}.

\begin{verbatim}
make_folds <- function(n, nfolds) {
  nn <- round(n / nfolds)
  sizes <- rep(nn, nfolds)
  sizes[nfolds] <- sizes[nfolds] + n - nn * nfolds
  b <- c(0, cumsum(sizes))
  ii <- sample(n)
  folds <- list()
  for (i in seq(nfolds))
    folds[[i]] <- ii[seq(b[i] + 1, b[i + 1])]
  folds
}

cv <- new_method_extension("cv", "cross validated",
                           method_extension = function(model, draw, out,
                                                       base_method) {
                             nfolds <- 5
                             err <- matrix(NA, ncol(out$beta), nfolds)
                             ii <- make_folds(model$n, nfolds)
                             for (i in seq_along(ii)) {
                               train <- model
                               train@params$x <- model@params$x[-ii[[i]], ]
                               train@params$n <- model@params$x[-ii[[i]], ]
                               train_draw <- draw[-ii[[i]]]

                               test <- model
                               test@params$x <- model@params$x[ii[[i]], ]
                               test@params$n <- model@params$x[ii[[i]], ]
                               test_draw <- draw[ii[[i]]]
                               fit <- base_method@method(model = train,
                                                         draw = train_draw,
                                                         lambda = out$lambda)
                               yhat <- test$x %*% fit$beta
                               ll <- seq(ncol(yhat))
                               err[ll, i] <- colMeans((yhat - test_draw)^2)
                             }
                             m <- rowMeans(err)
                             se <- apply(err, 1, sd) / sqrt(nfolds)
                             imin <- which.min(m)
                             ioneserule <- max(which(m <= m[imin] + se[imin]))
                             list(err = err, m = m, se = se, imin = imin,
                                  ioneserule = ioneserule,
                                  beta = out$beta[, imin],
                                  yhat = model$x %*% out$beta[, imin])
                           })
\end{verbatim}

Of course, if we later added \texttt{mcp} to the simulation, we could
easily incorporate cross-validation into it as well witih
\texttt{mcp\ +\ cv}.

We will typically put the code above in the file named
\texttt{method\_functions.R}.

\end{document}